# End-to-End Differentiable Photon Counting CT

Sen Wang, Yirong Yang, Jooho Lee, Grant M. Stevens, Adam S. Wang

***Abstract*— Quantitative imaging is an important feature of spectral X-ray and CT systems, especially photon-counting CT (PCCT) imaging systems, which is achieved through material decomposition (MD) using spectral measurements. In this work, we present a novel framework that makes the PCCT imaging chain end-to-end differentiable (differentiable PCCT), with which we can leverage quantitative information in the image domain to enable cross-domain learning and optimization for upstream models. Specifically, the material decomposition from maximum-likelihood estimation (MLE) was made differentiable based on the Implicit Function Theorem and inserted as a layer into the imaging chain for end-to-end optimization. This framework allows for an automatic and adaptive solution of a wide range of imaging tasks, ultimately achieving quantitative imaging through computation rather than manual intervention. The end-to-end training mechanism effectively avoids the need for direct-domain training or supervision from intermediate references as models are trained using quantitative images. We demonstrate its applicability in two representative tasks: correcting detector energy bin drift and training an object scatter correction network using cross-domain reference from quantitative material images.**

## I. INTRODUCTION

With the spectral measurements offered by spectral X-ray and CT imaging systems, the material decomposition (MD) method [1] maps X-ray measurements into basis material thicknesses, effectively decoupling the spectral-spatial terms from a polychromatic X-ray source, alleviating issues like beam hardening and metal artifacts and improving imaging quality. Dual-energy (DE) CT systems based on energy integrating detectors were first utilized to acquire the desired spectral measurements. More recently, photon counting detectors (PCDs) are being used to count individual photons and offer richer spectral information for material decomposition and quantitative imaging, with improved spatial resolution [2], [3], [4] and signal-to-noise ratio [5].

Accurate and robust physical calibration and correction is critical to ensure quantitative images. Conventionally, we establish references via physical measurements from dedicated experimental acquisitions or Monte Carlo simulations. For example, PCD models typically require calibration using a quasi-ideal X-ray source (e.g., a synchrotron source) [6] or semi-empirical slab measurements [7], [8]. Moreover, system-specific processing is generally needed since the operating conditions vary from system to system, demanding considerable human effort and time.

By viewing the entire processing chain as a whole, the end output of the aforementioned process is quantitative images. When errors exist in system calibration or correction within the imaging chain, they propagate through the system and are reflected as artifacts or bias in reconstructed quantitative images. If errors in the quantitative images can be determined, they could be used to update the imaging chain. This motivates the proposed differentiable PCCT imaging chain in this work, which can backpropagate loss defined in the quantitative image domain to the upstream physical measurement domain, thereby enabling cross-domain learning and optimization.

There have been several publications discussing the theory and tools to make CT reconstruction based on the linear Radon transform differentiable [9], [10], [11], enabling dual-domain learning across the sinogram and reconstructed image domains. Several studies have validated the corresponding benefits of the multi-domain scheme in training or optimizing reconstruction-related parameters and models, such as limited-angle reconstruction problem [9], [12], rigid-body motion calibration [13], [14], [15], and metal artifacts suppression [16].

Our focus in this work is to make the more upstream non-linear material decomposition process differentiable, in the spirit of differentiable optimization [17], [18]. As the conceptual pipeline in Fig. 1 shows, the proposed method extends existing differentiable reconstruction methods and further bridges the physical measurement (spectral counts) domain and the material sinogram domain, making the entire spectral imaging chain end-to-end differentiable.

Compared with conventional X-ray imaging system calibration or correction methods, which usually operate in isolated domains such as the physical measurement domain, the end-to-end differentiable method enables training of models used in the physical measurement domain with a loss defined in the quantitative image domain, while theoretically achieving equivalent convergence. As a result, this shifts the training reference to measurable and consistent quantitative images, making the method more robust and adaptable. We have shown the preliminary results in a simple empirical model correction [19].

This work is structured as follows: Section II introduces the proposed method, which streamlines the end-to-end differentiable PCCT imaging chain. Section III describes its evaluation tests in energy bin drift correction and deep scatter

Manuscript submitted for review January 17, 2026. This work was supported by GE HealthCare. (Corresponding author: Sen Wang.)

Sen Wang is with the Department of Radiology, Stanford University, Stanford, CA USA 94305. (email: senwang@stanford.edu)

Yirong Yang and Adam S. Wang are with the Department of Radiology and the Department of Electrical Engineering, Stanford University, Stanford, CA USA 94305. (e-mail: {yryangd, adamwang}@stanford.edu). Jooho Lee (e-mail: jooholee@yonsei.ac.kr) was with the Department of Radiology, Stanford University, Stanford, CA USA 94305 during his visit from Yonsei University, 50 Yonsei-ro, Seodaemun-gu, Seoul 03722, Republic of Korea.

Grant M. Stevens is with GE HealthCare, Waukesha, WI USA 53188. (e-mail: grant.stevens@gehealthcare.com)

correction. Section IV presents the results to reach the discussion and conclusion in Section V and Section VI, respectively.

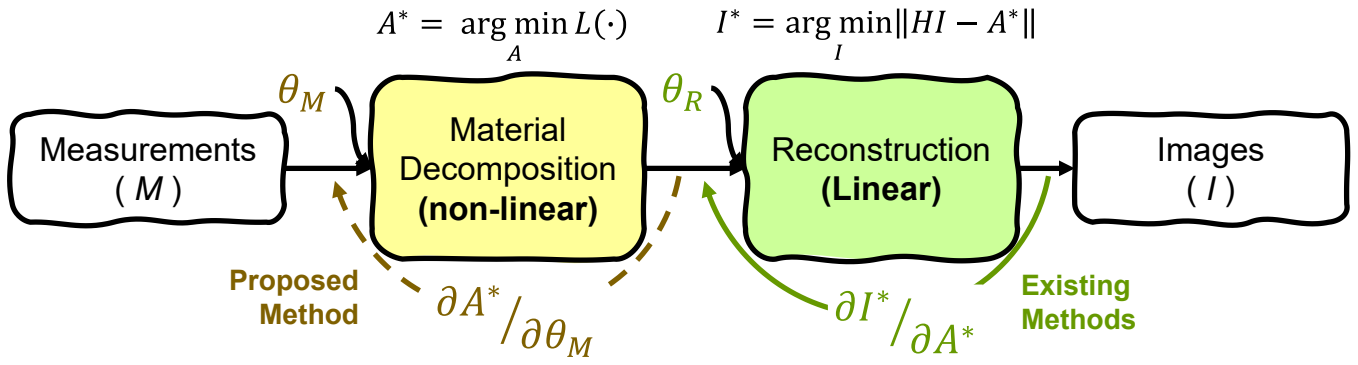

Fig. 1. Conceptual drawing of the proposed differentiable photon counting imaging chain. Many existing efforts have made reconstruction differentiable (usually based on Radon transform, a linear process), so that losses can be propagated between reconstructed images and sinograms to train parameters involved in reconstruction (i.e. $\theta_R$). Our focus is to make the upstream material decomposition differentiable (a non-linear estimation process) based on differentiable optimization theory [17], which enables the training of upstream parameters ($\theta_M$).

## II. Theory and Method

### A. Photon Counting CT Imaging Chain

We illustrate the general PCCT imaging pipeline in Fig. 2. The input measurements go through upstream calibration and processing that models or corrects non-ideal effects such as detector response, charge sharing, pulse pileup, and object scatter, which are in the physical measurement domain. Then, material decomposition is performed to map the physical measurements to quantitative basis material line integrals. Conventionally, the upstream calibration and processing models depend on underlying calibrations from manual measurements or Monte Carlo simulations, which can be tedious and time-consuming. For example, a semi-empirical forward model depends on slab measurements that can involve dozens of different combinations of basis materials and long acquisition time to suppress statistical noise [7], [8].

Fig. 2 also demonstrates the proposed end-to-end differentiable PCCT imaging chain that consists of both the forward path that generates quantitative MD projections/CT images from input measurements and the backward path that backpropagates the gradients from quantitative image domain to physical measurement domain. In the physical measurement domain, correction or calibration models usually yield explicit forms such as polynomials or neural networks [7], [20], [21], which are inherently differentiable. We categorize basis material projections and basis material images in the quantitative image domain for conciseness as they both have quantitative meaning and existing works have made image reconstruction differentiable [9], [10].

Material decomposition (in the projection domain) bridges the physical measurement domain and quantitative image domain. For dual energy or two energy bins, the mapping is deterministic and straightforward. However, PCDs can offer more than two energy bins, in which case maximum likelihood estimation (MLE) gives asymptotically efficient estimates of material thickness from spectral measurements [6]. MLE is a non-linear process, usually solved iteratively, thus it is not directly differentiable. In this work, we apply the theory from differentiable optimization [17] to make it differentiable and will introduce more details in the following section. We note that there are mainly two categories of material decomposition for quantitative X-ray CT imaging: projection-domain methods and image-domain methods. The focus of this work is projection-domain material decomposition, and the general idea from differentiable optimization can be readily applicable to image-domain or other methods.

### B. Differentiable Material Decomposition

In PCCT, PCDs count incident photons independently and record them to corresponding energy bins. The measurements depend on the source spectrum, the material attenuation, and the detector response as shown in equation (1), where $\lambda_i$ denotes the expected measurements (clean measurements) in the $i$-th energy bin, $D_i$ and $D_{i+1}$ are the lower and higher energy thresholds. $S_0(E')$, $\mu(x, E')$, and $h(E; E')$ denote the source spectrum, material attenuation coefficient, and detector response [6], respectively.

$$\lambda_i = \int_{D_i}^{D_{i+1}} \left( \int_0^\infty S_0(E') \exp\left( - \int \mu(x, E') dx \right) h(E; E')\, dE' \right) dE \quad (1)$$

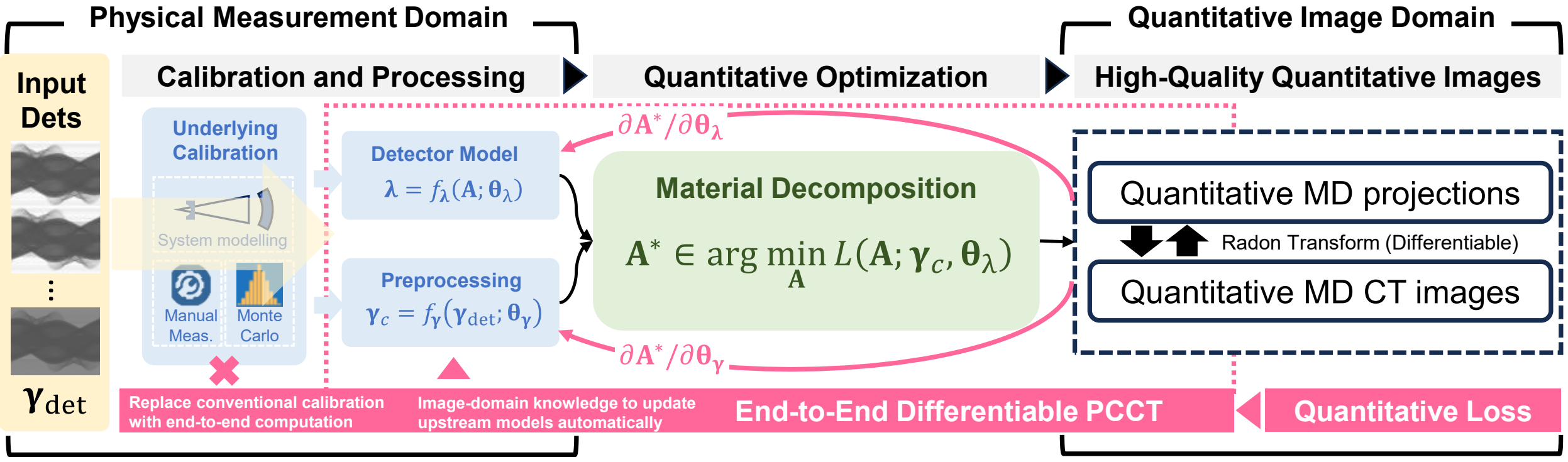

Fig. 2. Differentiable photon counting CT imaging pipeline, consisting of the forward path that processes input detected signal to generate quantitative material decomposition (MD) projections or reconstructed CT images, and the backward path that propagates quantitative loss defined in the quantitative image domain to the upstream physical measurement domain. The critical path is the introduction of differentiable optimization to the maximum likelihood estimation-based material decomposition in the imaging chain.

The material attenuation coefficients can be linearly represented using basis functions or basis materials $\mu(E) = \sum_k a_k f_k(E)$, where $f_k(E)$ are the basis functions or basis materials, $a_k$ are the decomposition coefficients, and $k$ means the $k$-th basis material, with which the photon transmission can be converted to equation (2):

$$\lambda_i = \int_{D_i}^{D_{i+1}} \left( \int_0^\infty S_0(E') \exp\left( -\sum\nolimits_k A_k f_k \right) h(E; E')\, dE' \right) dE \qquad (2)$$

where $A_k = \int a_k(x)dx$ denotes the line integral of decomposition coefficients that decouples the energy term. In diagnostic energies, the interactions between photons and materials mainly include the photoelectric effect and Compton scattering, meaning that $\mu(E)$ can be represented by two basis functions or basis materials if no K-edge materials exist. When there are K-edge materials, they can be added as additional basis materials.

The photon counting measurements are assumed to follow a Poisson distribution, meaning that $\gamma_i \sim \text{Poisson}(\lambda_i)$ is the noisy measurement in the $i$-th energy bin. Material decomposition in the projection domain estimates $\mathbf{A}$ from $\boldsymbol{\gamma}$. $\mathbf{A} = [A_k]_{1\times K}$ denotes the vector of basis material line integrals and $\boldsymbol{\gamma} = [\gamma_i]_{1\times M}$ denotes a vector representing the measurements in all energy bins, where $K$ and $M$ denote the number of basis materials and the number of spectral measurements, respectively. When $M > K$, the spectral information can be optimally utilized through maximum-likelihood estimation (MLE):

$$\mathbf{A}^* \in \arg\min_{\mathbf{A}} L_{MD}(\mathbf{A}; \boldsymbol{\gamma}), \qquad (3)$$
$$L_{MD}(\mathbf{A}) = \sum\nolimits_i (\lambda_i - \gamma_i \log(\lambda_i)).$$

Consider the common PCCT imaging chain as shown in Fig. 2, where the PCD model is calibrated to represent $\lambda_i$ and a series of preprocessing steps is applied onto raw measurements ($\boldsymbol{\gamma}_{det}$) before material decomposition, as denoted by $\boldsymbol{\lambda} = f_\lambda(\mathbf{A}; \boldsymbol{\theta}_\lambda)$ and $\boldsymbol{\gamma}_c = f_\gamma(\boldsymbol{\gamma}_{det}; \boldsymbol{\theta}_\gamma)$, respectively. The material decomposition process depends on parameters from both $\boldsymbol{\theta}_\lambda$ and $\boldsymbol{\theta}_\gamma$. This leads to an expanded formulation of MLE-based material decomposition that explicitly incorporates the upstream parameters involved in the process:

$$\mathbf{A}^* \in \arg\min_{\mathbf{A}} \sum\nolimits_i \Big( f_{\lambda_i}(\mathbf{A}; \boldsymbol{\theta}_\lambda) - f_{\gamma_i}(\boldsymbol{\gamma}_{det}; \boldsymbol{\theta}_\gamma) \log\big( f_{\lambda_i}(\mathbf{A}; \boldsymbol{\theta}_\lambda) \big) \Big) \qquad (4)$$

The idea of differentiable material decomposition is to find the gradients that relate $\mathbf{A}^*$ to $\boldsymbol{\theta}_\lambda$ and $\boldsymbol{\theta}_\gamma$, i.e., $\partial \mathbf{A}^* / \boldsymbol{\theta}_\lambda$ and $\partial \mathbf{A}^* / \boldsymbol{\theta}_\gamma$, which falls within the concept of differentiable optimization [17], [22].

The MLE problem in equation (3) and (4) is usually solved iteratively to find its optimal solution that follows the partial derivative condition $\frac{\partial L}{\partial \mathbf{A}}|_{\mathbf{A}^*} = 0$. We can obtain its general-form gradient based on the Implicit Function Theorem [17], [22]:

$$\frac{\partial \mathbf{A}^*}{\partial \boldsymbol{\theta}} = -\left( \frac{\partial^2 L_{\text{MD}}}{\partial \mathbf{A}^2} \right)^{-1} \frac{\partial^2 L_{\text{MD}}}{\partial \boldsymbol{\theta} \partial \mathbf{A}}, \qquad (5)$$

where $\boldsymbol{\theta}$ represents the parameters associated with the optimization, $\boldsymbol{\theta}_\lambda$ and $\boldsymbol{\theta}_\gamma$ in our case. As we have the explicit form of $L_{\text{MD}}$, $\left(\frac{\partial^2 L_{\text{MD}}}{\partial \mathbf{A}^2}\right)^{-1}$, $\frac{\partial^2 L_{\text{MD}}}{\partial \boldsymbol{\theta}_\lambda \partial \mathbf{A}}$, and $\frac{\partial^2 L_{\text{MD}}}{\partial \boldsymbol{\theta}_\gamma \partial \mathbf{A}}$ can be computed analytically using the chain rule, as shown in the Appendix.

Thus, the differentiable material decomposition makes the PCCT imaging chain end-to-end differentiable. Theoretically, it yields two properties: 1) It can bring loss back from quantitative material projections or reconstructed images for cross-domain/all-domain learning and optimization. 2) It is realized in tensor form, meaning that we can utilize the inherent similarity and correlation in quantitative projections or reconstructed images. Alternatively, the iterative estimation process can be made differentiable by unrolling the solver [23]. However, it is very GPU-demanding and becomes quite impractical for PCCT which contains thousands of detector pixels. Additionally, in practical detectors, each pixel may have its own detector model or corrections. A brief comparison is included in the Appendix. Equation (4) also shows that in each iteration we need to solve the forward problem first to obtain the estimated material thickness $\mathbf{A}^*$ and then calculate the gradients at the solution for the backward propagation. The forward problem is in principle solver-agnostic. In this work, we generally used a parallelized Newton-Raphson's method implemented on GPUs.

### *C. From a Perspective of Bilevel Optimization*

The differentiable PCCT image chain and the enabled cross-domain optimization tasks can be viewed more generally as bilevel optimization [24], [25], where the outer objective is defined in terms of the optimal solution to an inner objective:

$$\boldsymbol{\theta}^* \in \arg\min_{\boldsymbol{\theta}} \mathcal{J}_{\text{out}}(\boldsymbol{\theta}, \mathbf{A}^*) \qquad (6)$$
$$\text{s.t.} \quad \mathbf{A}^* \in \arg\min_{\mathbf{A}} \mathcal{J}_{\text{in}}(\boldsymbol{\theta}, \mathbf{A}).$$

In the context of PCCT imaging, the inner optimization $\mathcal{J}_{\text{in}}$ is MLE for material decomposition and the outer optimization $\mathcal{J}_{\text{out}}$ can be defined in quantitative projection or image domains. $\boldsymbol{\theta}$ can be the parameters of interest along the imaging chain.

TABLE I
OVERVIEW OF TESTS IN DIFFERENT UPSTREAM DOMAINS

| | | Upstream Domains | |
|---|---|---|---|
| | | Model domain | Counts domain |
| Model complexity | Empirical model | Test 1: Detector energy threshold correction | \ |
| | Deep network | \ | Test 2: Scatter correction network |

## III. EVALUATION

As illustrated in Fig. 2, there are two main domains upstream to material decomposition: the PCD counts domain where we usually carry out a series of preprocessing steps, and the PCD model domain that contains parameterized detector models. Aiming to evaluate the effectiveness of the proposed approach, we conducted tests of varying complexity in both domains as listed in TABLE I.

### A. Sanity Check of the Differentiable PCCT Imaging Chain: Perturbation Analysis

The gradients from equation (5) can be validated by manually applying perturbation to the measurements or model parameters as a sanity check, i.e., applying $\Delta\boldsymbol{\theta}$ onto $\boldsymbol{\theta}$ to check the alignment between corresponding $\Delta\mathbf{A}^*$ and $(\partial\mathbf{A}^*/\partial\boldsymbol{\theta})\cdot\Delta\boldsymbol{\theta}$. We carried out the perturbation test for material decomposition around water=20 cm and Ca=1 cm. Newton's method was used as the solver. We used a 120 kV spectrum, two energy bins (with thresholds 15, 50 keV), and a realistic spectral response of a 60-mm-thick Si detector [26] in the test. A perturbation of $\pm1\%$ was applied in counts domain measurements, and a $\pm2$ keV perturbation was applied onto detector energy thresholds.

### B. Cross-Domain PCD Model Energy Threshold Error Correction

In this test, we aim to evaluate if the differentiable PCCT imaging chain can correct an error in the energy threshold of the PCD model using quantitative material decomposition references in the projection domain or image domain. We used the generic PCD model as shown in equation (2) as an example and the parameter of interest ($\boldsymbol{\theta}_{\boldsymbol{\gamma}}$ in equation (4)) is now the threshold $D_i$ in equation (2). By leveraging $\partial\int_0^{D_i} s(E)dE/\partial D_i = s(D_i)$ and applying the chain rule, we can obtain the analytical form of $\partial\mathbf{A}^*/\,\partial D_i$.

Currently, a common way to calibrate PCD models is to carry out slab measurements with different thicknesses and material combinations. The calibration can be tedious and time-consuming. In this test, we aim to show that with the proposed framework, we can realize the PCD model correction using references in the form of either material thickness or material images, thus providing improved flexibility with scan-based calibration and correction. We carried out numerical simulation tests and used a realistic spectral response from Monte Carlo simulation [26] and a 120 kV source spectrum from CatSim [27]. The same realistic spectral response of a 60-mm-thick Si detector [26] was used.

#### 1) Material thickness reference in projection domain

In the material thickness-based correction and calibration, we used a series of basis material combinations of different thicknesses. Water and calcium were selected as the basis materials (water: 0:2:30 cm, calcium: 0:0.2:2 cm). Without lack of generalizability, we used a hypothetical detector with four energy bins and energy thresholds 15, 33, 45, 59, and 120 keV. A fixed bias of 0.5 keV was applied to the detector energy thresholds, and consequently, bias was introduced in the material decomposition. We used the differentiable framework to minimize the material decomposition thickness error by updating the bin thresholds as follows:

$$\Delta\mathbf{D}^* = \arg\min_{\Delta\mathbf{D}} \left\| \arg\min_{\mathbf{A}} L_{MD}(\mathbf{A};\, \boldsymbol{\gamma}, \mathbf{D}_b - \Delta\mathbf{D}) - \mathbf{A}_{ref} \right\|, \quad (7)$$

where $\mathbf{D}_b$ means the detector energy bin with bias, $\Delta\mathbf{D}^*$ is the threshold correction, and $\mathbf{A}_{ref}$ is the reference material decomposition, assumed to be known *a priori*. $L_{MD}(\cdot)$ means the log-likelihood function used in material decomposition as shown in equation (3).

#### 2) Material image-based PCD model calibration with differentiable PCCT

Material thickness-based calibration usually requires manually placing slabs to acquire reference thickness combinations. The differentiable framework can also enable image-based calibration, where we can easily establish the reference using known phantoms to simplify the calibration to one or a few phantom scans. Four detector energy bins were set to the same thresholds of 15, 33, 45, 59, and 120 keV. Given high-quality image references, we can use the differentiable pipeline to calibrate the PCD model by minimizing errors or artifacts in reconstructed images as follows:

$$\Delta\mathbf{D}^* = \arg\min_{\Delta\mathbf{D}} \left\| f_{\text{iRadon}}\left( \arg\min_{\mathbf{A}} L_{MD}(\mathbf{A};\, \boldsymbol{\gamma}, \mathbf{D}_b - \Delta\mathbf{D}) \right) - f_{\text{iRadon}}\left(\mathbf{A}_{ref}\right) \right\|, \quad (8)$$

where $f_{\text{iRadon}}(\cdot)$ denotes the image reconstruction operator. For simplicity and without loss of generality, we simulated a parallel-beam geometry using the LEAP CT toolbox [14] with off-the-shelf differentiable reconstruction. We used a detector array of 1×256 pixels (each 0.15 cm wide) and manually applied 0.5 keV bias onto the 75th and 90th detector pixels, showing the capability to correct multiple bins and pixels simultaneously. A numerical phantom consisting of an elliptic water region and two calcium regions was used in the detector energy bin calibration, as shown in Fig. 3.

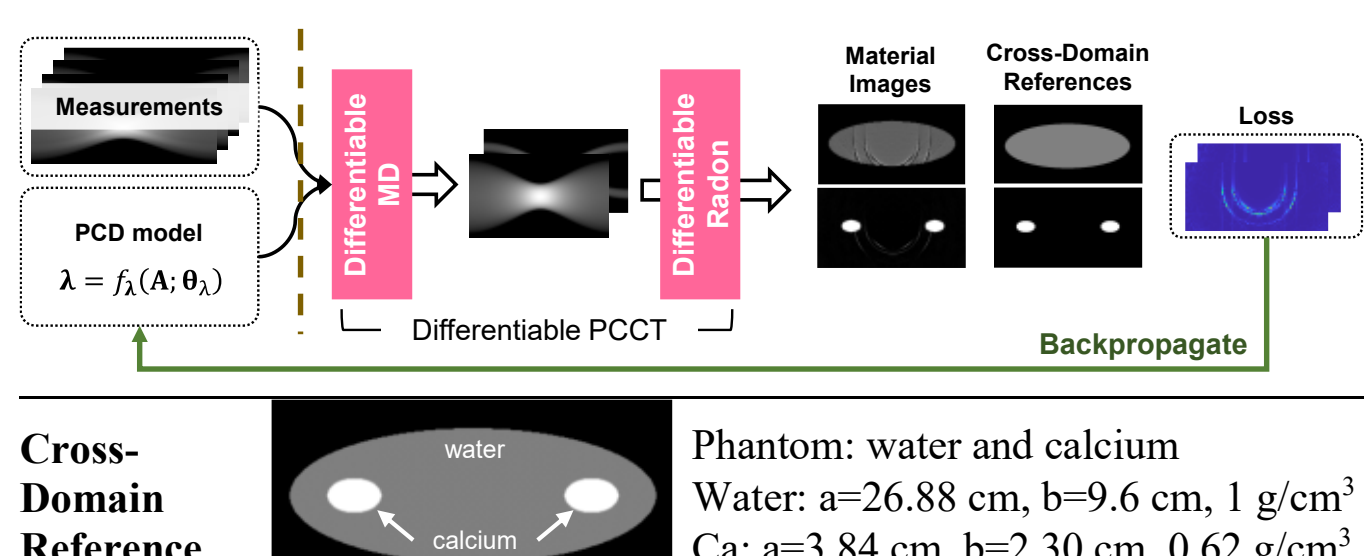

Fig. 3. Image-based calibration using the differentiable photon counting CT imaging chain. Energy threshold errors will cause ring artifacts in the reconstructed image. With known phantoms, we can minimize the difference between the reconstructed image and ground-truth reference to correct the energy threshold error in a cross-domain manner using the differentiable chain. An elliptic phantom of water and calcium was used in the simulation study, where *a* and *b* denote the width and height of the respective ellipses.

We purposely kept the phantom non-isotropic so that each pixel detects different combinations of material thicknesses across 180 different angles during gantry rotation. Noiseless projections were acquired during the calibration, which emulates high-dose or repeated scans in practice. After the phantom-based calibration, we applied the corrected thresholds to a simulated PCCT scan of an abdomen, using an axial image from the KiTS21 dataset [28]. This simulates a scenario of phantom-based calibration followed by clinical scans.

### C. Cross-Domain Deep Scatter Correction

We present another application of the differentiable PCCT imaging chain to train a neural network for the object/patient scatter correction shown in Fig. 4. Among various existing methods for scatter correction, neural network-based methods [20], [21], [29] have shown stronger effectiveness than conventional methods (e.g., kernel-based methods [29]). We demonstrate that it is possible to train the network in a cross-domain manner using the differentiable imaging chain without direct scatter references, thereby eliminating the complex and time-consuming Monte Carlo simulation for generating scatter references [20].

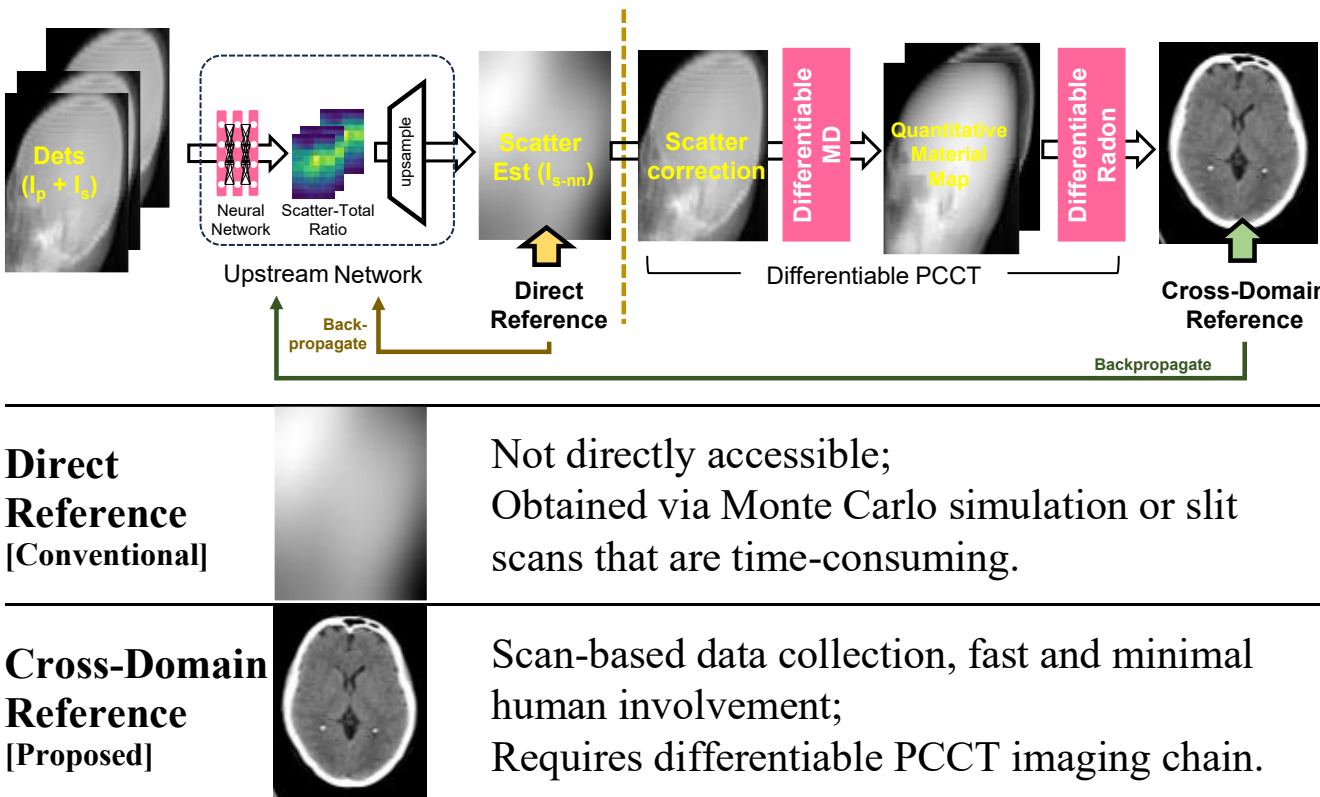

Fig. 4. Object scatter correction using the differentiable photon counting CT imaging chain. With the differentiable imaging chain, we can backpropagate loss defined in the quantitative image domain to train models. In this case, we use quantitative reconstructed images as reference (cross-domain) to train the model for scatter correction in the physical measurement domain without direct scatter references.

In this test, we built a numerical simulation dataset for the evaluation using 50 cases from the Stanford AIMI SinoCT dataset [30]. For training purposes, 48 projections were acquired over a 240-degree rotation with a step size of 5 degrees under a hypothetical cone-beam flat-detector imaging configuration with source-isocenter distance of 62.5 cm and source-detector distance of 109.7 cm. Primary projections were simulated using the LEAP CT toolbox [11]. Object/patient scatter was simulated through a GPU-enabled Monte Carlo tool [31] with a projection dimension of 384×256 and $10^9$ photons per projection from an isotropic 120 kV spectrum. The Monte Carlo tool was adapted to output photon counting measurements. In the test phase, we simulated patient projections covering 360 degrees with a step size of 1 degree. It took 20 seconds per view to finish the Monte Carlo simulation and approximately 2 hours for a 360-degree scan, comparable to [20], meaning that it took several days to build the dataset. This reflects the time-consuming nature of Monte Carlo reference generation, which we aim to avoid using the end-to-end differentiable chain. The magnitude of scatter was scaled to around 10% scatter-to-primary ratios (SPR). A realistic spectral response of a 60-mm-thick Si detector [26] was used to generate count measurements in four energy bins (thresholds = 15, 33, 45, 59, 120 keV). A 3×3 Gaussian filter was further applied onto scatter measurements to ensure the spatial smoothness of scatter intensity in each energy bin.

The total dataset was split into training (1440 projections from 30 patients), validation (480 projections from 10 patients), and test set (3600 projections from 10 patients). A convolutional neural network was used to gradually map the input detected signal (384×256×4) to the scatter-to-total counts ratio (12×8×4) that leverages the low spatial frequency property of scatter [20]. The network contains a 4-stage encoder network that gradually extracts features to 32×16×64 and a bottleneck layer maps the features to 12×8×4, following the concept in [20]. An additional Sigmoid layer was added as the final layer for normalization, ensuring the output remained within 0 to 1 (since scatter counts are a subset of total counts). Each encoder layer contains two 3×3 convolutional layers with BatchNorm and ReLU. The number of channels for the encoder layers is [64, 128, 128, 64], respectively. The bottleneck contains two 3×3 convolution layers, and the last convolutional layer has a stride of 2 to perform downsampling.

The models were trained for 200 epochs, minimizing L1 loss with Adam optimizer (initial learning rate = $10^{-3}$, 0.1 decay every 100 epochs) on our workstation (2 NVIDIA TITAN RTX GPUs, Intel Core i9-9960X CPU @ 3.10 GHz). Weights from the best validation were saved for the test. A parallelized Newton–Raphson method was used to solve the material decomposition problem.

Empirically, we found that, in the forward estimation step, imposing a bounding box on the decomposition results ($-5 \leq$ water (cm) $\leq 30$, $-5 \leq$ Ca (cm) $\leq 10$) and applying a gradient constraint of $5\times10^{-4}$ to the relative counts layer is necessary and effective for stabilizing the cross-domain optimization. In each epoch, we need to carry out material decomposition on all the cases, which takes about 400 seconds per epoch, or approximately 20 hours to complete the full training (200 epochs). In our previous work, we proposed a proxy method to accelerate the material decomposition process [32], which preserves the value and Jacobian of iterative MLE MD through derivative-aware training [25], [33]. We also showed their equivalence in the training that makes the entire training much faster (Appendix), allowing the training to be completed within approximately 3 hours.

In addition to the deep learning methods including conventional direct training and the proposed cross-domain training (illustrated in Fig. 4), we also included an empirical scatter kernel method for comparison [29].

## IV. Results

### A. Perturbation Analysis Result of Differentiable PCCT

The results of our perturbation test are shown in Fig. 5, including both counts perturbation in the counts domain and detector threshold perturbation in the PCD model domain. Results from the material decomposition solver are the grid plots with certain curvature, which also reflects the nonlinearity of the process. Local gradients form the tangent plane of the curved surface as expected. This result validates the correct realization of the gradients in our pipeline, which is the foundation of the following cross-domain learning and optimization.

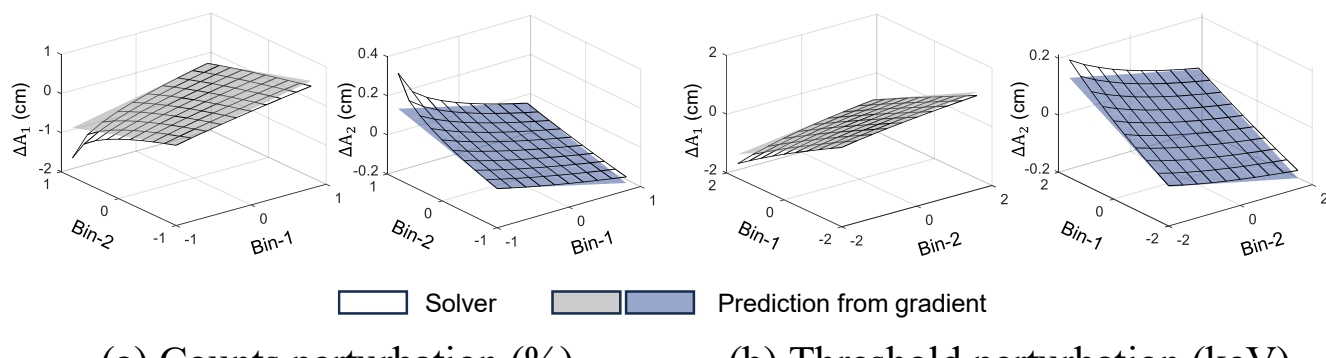

(a) Counts perturbation (%) (b) Threshold perturbation (keV)

Fig. 5. Perturbation test to validate the gradient calculation against the reference solver for water ($A_1$) and calcium ($A_2$) due to perturbations in (a) counts (%) and (b) bin thresholds (keV).

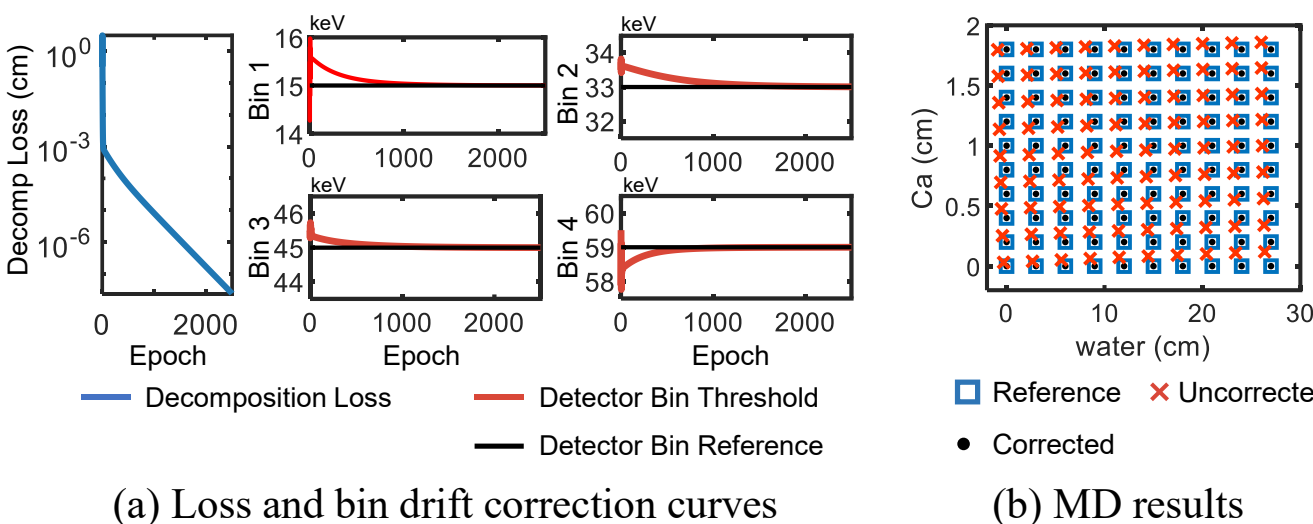

(a) Loss and bin drift correction curves (b) MD results

Fig. 6. Correction of detector energy threshold errors in the photon counting detector model with material thickness reference. (a) Using the proposed differentiable material decomposition layer, we can correct bin drift errors by minimizing the loss in decomposition. (b) After correction, material decomposition errors are eliminated.

## B. Photon Counting Detector Bin Drift Correction using Differentiable PCCT

The results from material thickness-based calibration are shown in Fig. 6, which corrects the energy threshold error in one pixel using different basis material combinations. Fig. 6 (a) shows by minimizing the material decomposition loss, the threshold error gradually converges to zero, demonstrating the effectiveness of the cross-domain optimization. The material decomposition results are illustrated in Fig. 6 (b), where blue squares denote the ground-truth reference, red crosses are the decomposition results when bias exists in detector energy thresholds (detector bin drift), and black dots are the corrected material decomposition after the threshold error correction through differentiable PCCT.

Material decomposition errors caused by inaccurate PCD model parameters in individual detectors will lead to ring artifacts in the reconstructed phantom images as shown in Fig. 7. We also show that with the known ground-truth references we can calculate the difference image as the loss. By minimizing the loss, we eventually corrected the threshold errors in multiple pixels simultaneously and eliminated the ring artifacts. The left column shows the reference images of water and calcium, and the middle column shows images from biased detector bin thresholds, where ring artifacts are evident. The right column shows images after correction, where the ring artifacts are effectively eliminated.

As mentioned, for conventional energy threshold calibrations, the required material combinations are obtained by slab measurements, where different combinations of basis materials are placed and scanned manually. This test shows that such correction can be realized through phantom scan-based calibration using the differentiable PCCT chain, which implicitly takes advantage of all the combinations of materials from different projection angles, with a potential to reduce the amount of human effort in the calibration phase.

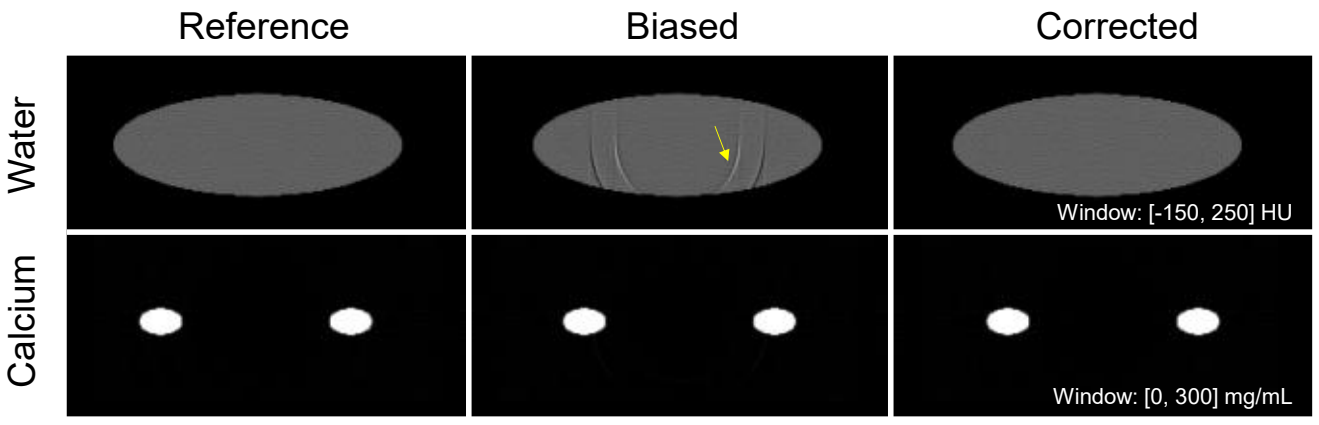

Fig. 7. Calibration and correction of detector energy thresholds in the PCD model with material image reference. (Left) reference water and calcium images, (middle) reconstruction with biased bin thresholds, and (right) reconstruction after bias correction.

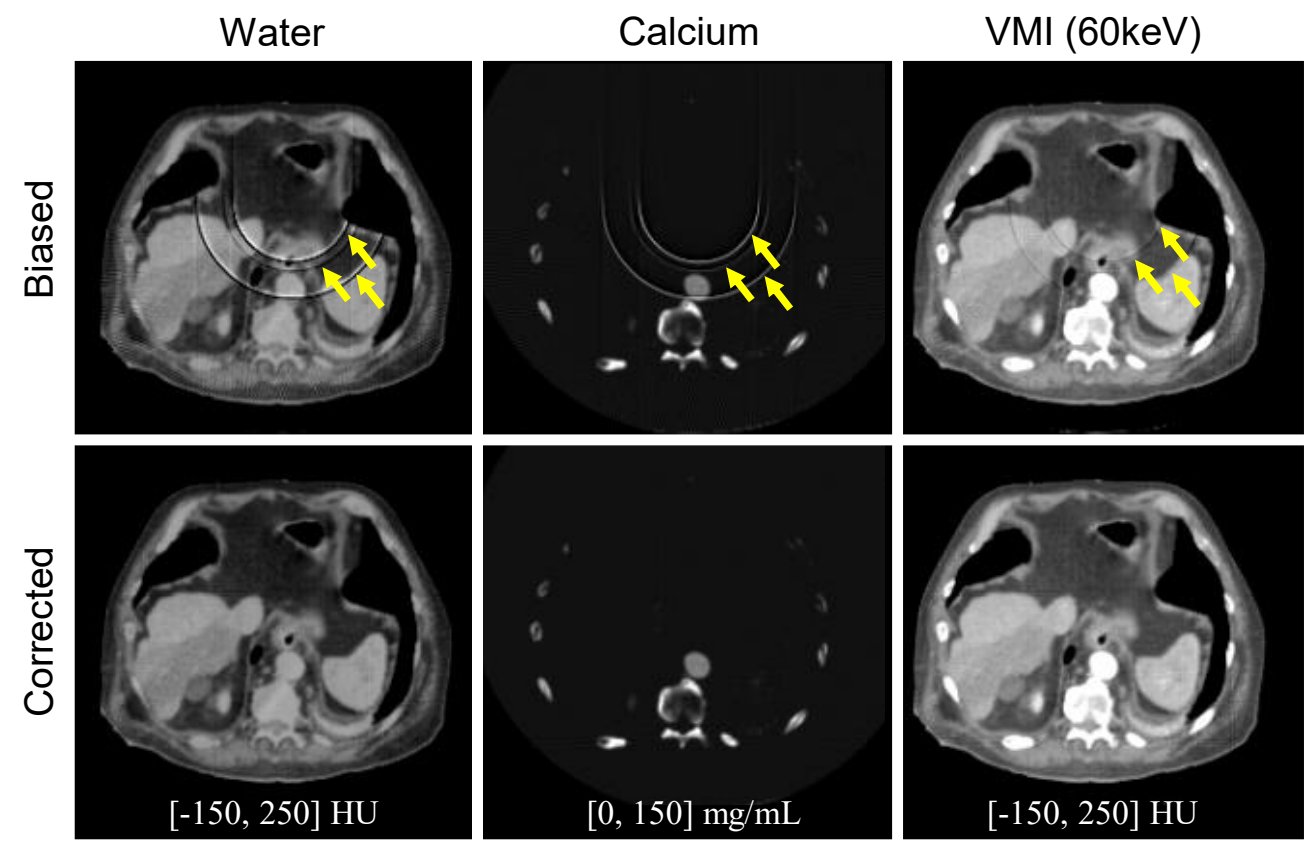

Fig. 8. Material decomposition after the scan-based phantom calibration. (Top) Biased thresholds yield rings in water, calcium, and virtual monoenergetic (60 keV) images. (Bottom) Phantom-based calibration also corrects rings in subsequent scans.

Once the detector bin drift is fixed in the phantom scan, we can expect material decompositions of other objects to be accurate so that the reconstructed images are free of ring artifacts. This is illustrated in Fig. 8 for the biased and subsequent corrected thresholds on the simulated abdominal PCCT scan.

## C. Scatter Correction Test Results

We evaluated the test performance of both conventional kernel-based and neural network-based scatter correction methods in both material projection domain and reconstructed image domain as listed in TABLE II and illustrated representative basis images and virtual monoenergetic images (VMIs) at 60 keV from the test set in Fig. 9 with RMSE labelled in each figure.

For neural network-based methods, Direct NN and Cross NN denotes the model trained with direct scatter loss and with cross-domain image loss, respectively. Overall, the neural network methods performed significantly better than conventional kernel-based methods, which aligns with previous findings [21]. The neural network trained in a cross-domain manner utilizing the differentiable PCCT chain achieved nearly equivalent performance with the neural network trained using direct scatter reference, demonstrating the image-domain loss is successfully backpropagated in the differentiable imaging chain to the upstream photon counting measurements domain to train the neural network. In Fig. 9, we find comparable performance between Cross NN and Direct NN, and both outperformed the conventional kernel-based method.

TABLE II
TEST PERFORMANCE OF SCATTER CORRECTION FROM DIFFERENT METHODS

| Evaluation Domain | Metrics | Uncorrected | Empirical Kernel | Neural Networks | |
|---|---|---|---|---|---|
| | | | | Direct NN [need intermediate reference] | Cross NN [end-to-end] |
| Projection | Water RMSE (cm) | 0.2587 | 0.1297±0.1512 | 0.0632±0.0322 | **0.0628±0.0341** |
| | Calcium RMSE (cm) | 0.0374 | 0.0332±0.0151 | 0.0146±0.0077 | **0.0113±0.0057** |
| | SPR (%) | 10.47% | 5.56±3.37% | 2.30±1.94% | **2.12±1.73%** |
| Reconstruction | Water RMSE (mg/mL) | 34.95 | 19.43±6.50 | 11.80±4.17 | **10.06±3.28** |
| | Calcium RMSE (mg/mL) | 17.10 | 6.06±1.41 | 5.51±2.63 | **4.38±2.15** |
| | VMI (60 keV) RMSE (HU) | 68.82 | 30.16±5.94 | 20.13±9.88 | **17.60±8.10** |

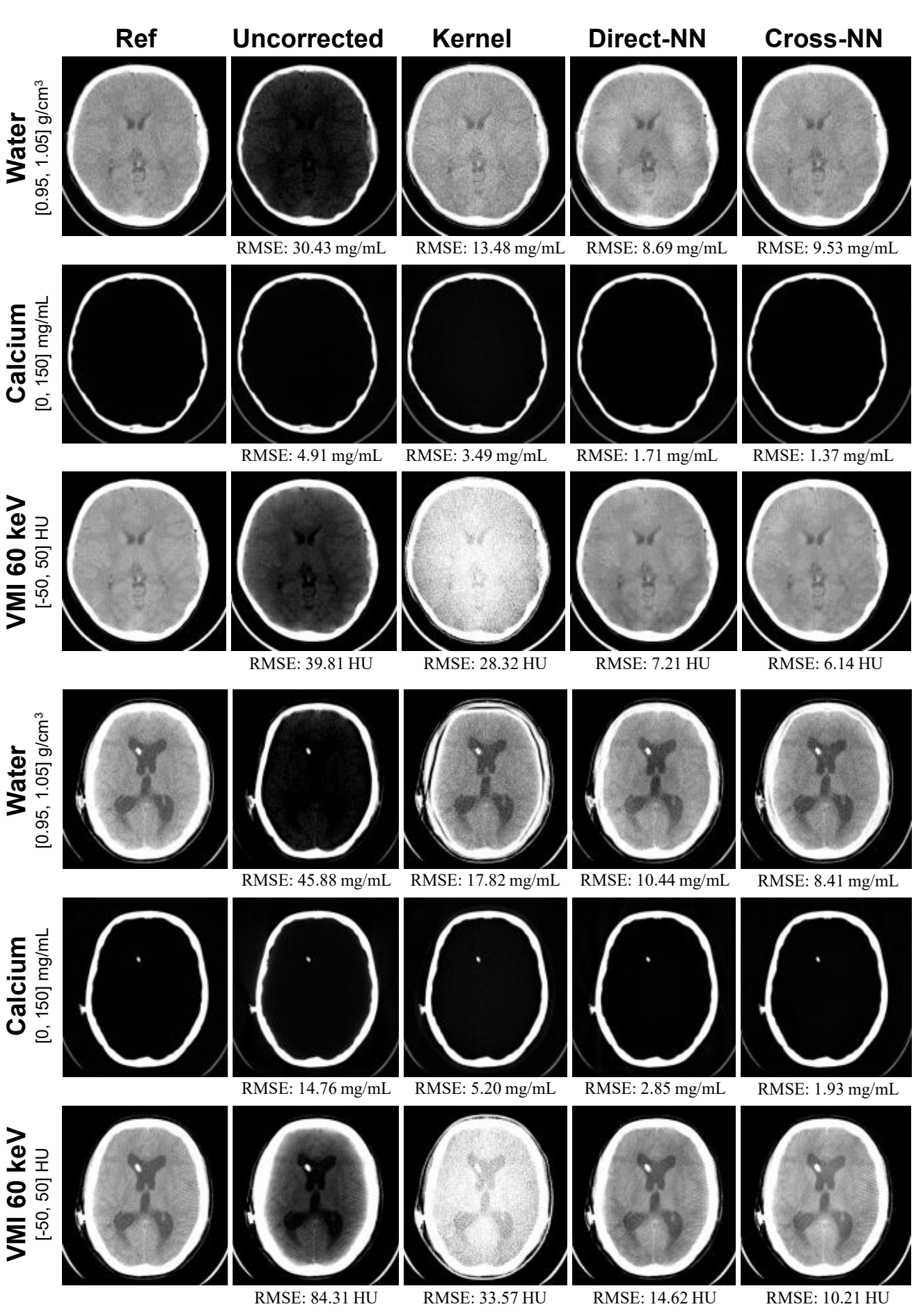

Fig. 9. Representative reconstructed images of 2 test cases. Water, calcium, and VMI at 60 keV are illustrated with corresponding RMSEs labelled on the image. Display windows are labelled as well.

Fig. 10 shows representative water and calcium decomposition results in the projection domain from the test set, which presents similar results to Fig. 9 and TABLE II in another domain.

Fig. 11 presents detailed statistics of the test performance of the methods listed in TABLE II, including SPR, water image RMSE, calcium image RMSE, and VMI (60 keV) RMSE, and demonstrates statistically equivalent performance between Direct NN and Cross NN. Based on the mean values, slight differences can be observed in the behavior of Cross NN and Direct NN. Cross NN shows a modest advantage in the RMSEs of reconstructed images, while Direct NN performs slightly better in SPR correction in the off-center region. This aligns with the training objectives of the networks: Cross NN is optimized to produce accurate reconstructed images, while Direct NN is trained to minimize scatter loss. In addition to not requiring direct scatter references, another advantage of end-to-end training appears to be more accurate quantitative imaging.

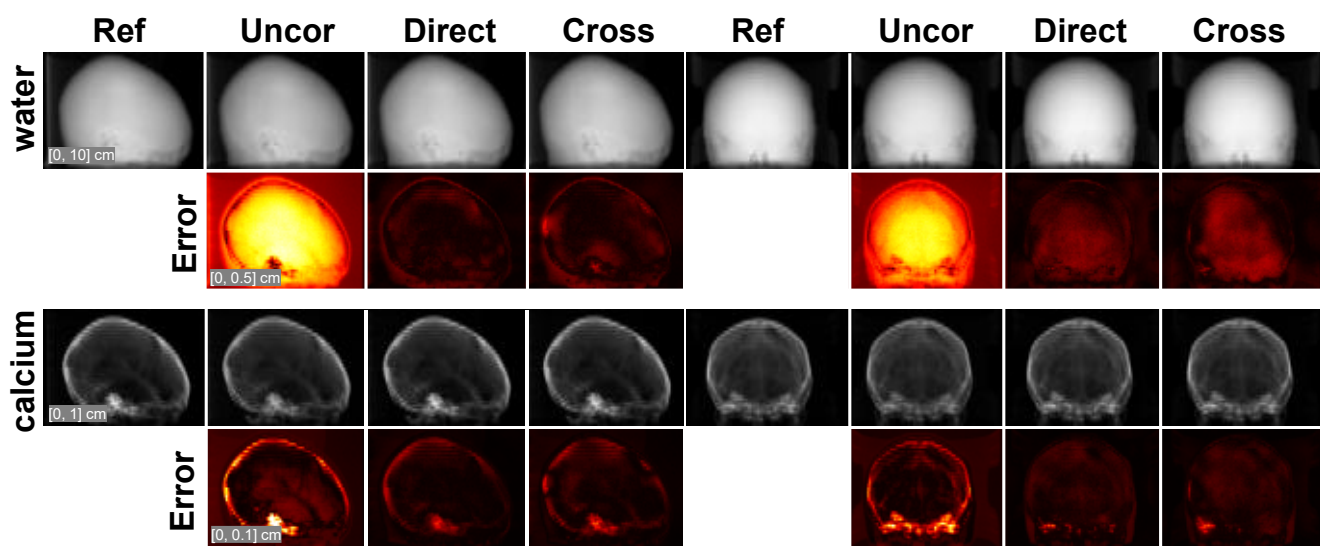

Fig. 10. Material decomposition results into water and calcium images of a case in the test dataset. Two representative views are illustrated in the left and right halves. Uncor denotes the decomposition results without scatter corrections, where strong bias exists. Networks trained using either cross-domain references (Cross) or direct scatter references (Direct) reduce decomposition errors.

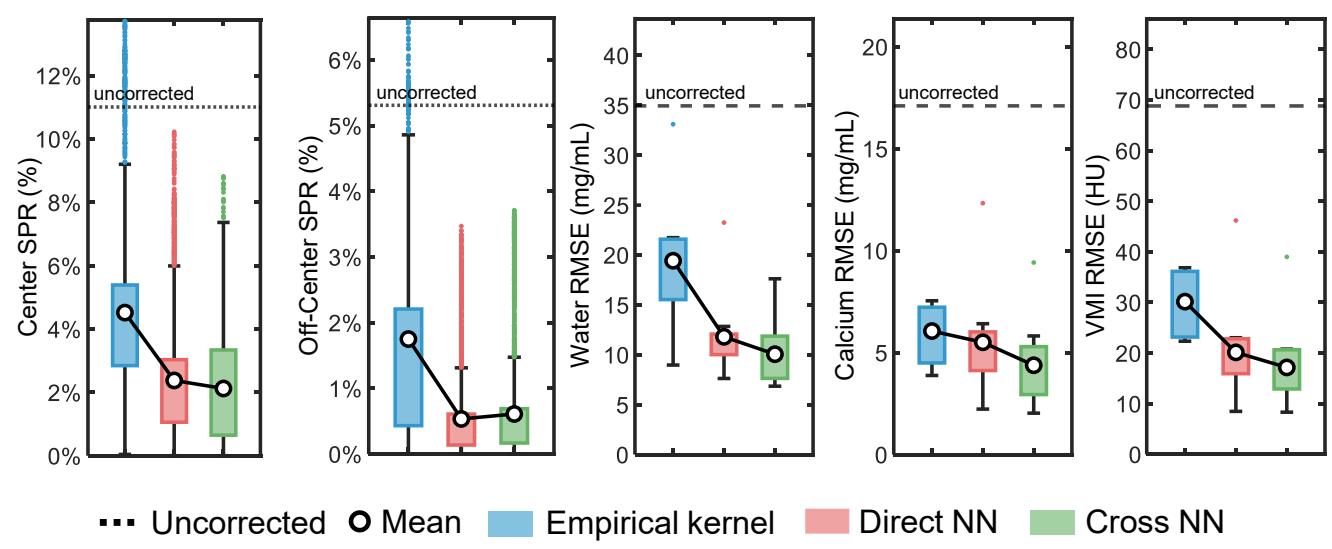

Fig. 11. Test performance of scatter correction models.

## V. DISCUSSION

In this work, we proposed a differentiable PCCT imaging pipeline, especially showing a method to make the iterative MLE-based non-linear material decomposition differentiable through differentiable optimization techniques. It connects all domains across the PCCT imaging chain and enables parameter training and learning. Our work is motivated by recent progress in quantitative spectral CT imaging and rapid development of AI. By making the PCCT imaging chain differentiable, we can intuitively define quantitative losses to train data-driven AI applications for upstream imaging tasks.

We demonstrated the potential use of the method in optimizing parameters in both lightweight empirical models for detector energy threshold correction and neural networks for object scatter estimation. From these examples, we found that the imaging chain successfully trained models using loss defined in quantitative material sinograms and reconstructed material images in a cross-domain manner, which shows promising capability to automate calibration or simplify training references, while achieving equivalent performance with models trained with direct reference. In the scatter correction case, we even observed improved performance in reconstructed images, where the training target was set and the cross-domain loss was defined, showing potential advantage of the end-to-end training.

In addition, the differentiable imaging chain can be readily applied to a broader set of applications, such as calibrating other parameters in PCD models [7], [34], pulse pileup correction [35], [36], etc. Future work investigating these problems can consider whether scenario-specific designs may be required, e.g., empirical models or neural network models. In general, the differentiable material decomposition bridges the entire PCCT imaging chain, allowing us to train and optimize models by distilling knowledge from the quantitative image domain. It is also possible to incorporate the differentiable imaging chain with other downstream tasks in computer-aided medical diagnostics, such as task-driven imaging optimization.

However, compared to direct training, maximum-likelihood estimation is performed during each epoch of cross-domain training as the references are the quantitative images, even when the inputs still contain errors. This makes numerical stability a potential concern for the method, which can be more thoroughly tested in future work. Instead, in our current implementation, we incorporated careful design choices to avoid breaking the chain: in detector energy threshold correction, we limited the error to 0.5 keV, where decomposition errors are moderate; in scatter correction network training, we constrained the network to learn the scatter-to-total ratio, which is known to be within (0, 1), ensuring that its output remains physically meaningful and does not break the material decomposition (for example, negative counts are avoided). Future work can further understand and discuss the corresponding numerical stability. Another interesting topic for discussion is the training efficiency of direct training versus cross-domain training. Material decomposition in multi-bin photon counting systems contracts measurements from multiple energy bins to two-channel or three-channel material images. It remains unclear how deterministic information propagates and backpropagates, and whether there is information loss in the decomposition that could lead to increased data demand in cross-domain training.

Moreover, in the current work, all training and optimization are based on some type of ground truth as the reference, i.e., supervised learning. Since image-domain references are not always available, we are interested in exploring the feasibility of introducing unsupervised/self-supervised learning strategies with the differentiable imaging chain. Also, in addition to MLE-based MD, there are several other empirical methods for material decomposition, such as polynomial-based look-up-table methods or linearized MD. These methods generally rely on a direct mapping from measurements to material thickness using explicit functions or operators; thus, they are inherently differentiable. Comparative studies between differentiable MLE-MD and these differentiable empirical MD methods can be conducted in the future.

## VI. Conclusion

In this manuscript, we present a method to make non-linear material decomposition differentiable in photon counting CT imaging, enabling the entire imaging chain to be differentiable when combined with existing differentiable reconstruction methods. We demonstrate its use in correcting detector threshold errors and object scatter correction. With the differentiable PCCT imaging chain, we can train and optimize models in a cross-domain manner, potentially facilitating a broader range of applications.

## Appendix

### A. Gradients of Differentiable Material Decomposition

In this section, we present the gradients shown in Fig. 2 and equation (5) that make the material decomposition differentiable. With the negative log-likelihood function in material decomposition $L_{MD} = \sum_i (\lambda_i - \gamma_i \log(\lambda_i))$ shown in equation (3), we first obtain the first-order derivative as $\partial L_{MD}/\partial A_k = \sum_i (1 - \gamma_i/\lambda_i) \cdot \partial\lambda_i/\partial A_k$. From this, the second-order derivatives can be written as follows:

$$\frac{\partial^2 L_{\mathrm{MD}}}{\partial \mathbf{A}^2} = \left[\frac{\partial^2 L_{\mathrm{MD}}}{\partial A_m \partial A_k}\right]_{K\times K} = \left[\sum_i \left(1 - \frac{\gamma_i}{\lambda_i}\right)\frac{\partial^2 \lambda_i}{\partial A_m \partial A_k} + \frac{\gamma_i}{\lambda_i^2}\frac{\partial \lambda_i}{\partial A_m}\frac{\partial \lambda_i}{\partial A_k}\right]_{K\times K} \tag{A-1}$$

$$\frac{\partial^2 L_{\mathrm{MD}}}{\partial \boldsymbol{\theta}_{\boldsymbol{\gamma}} \partial \mathbf{A}} = \left[\frac{\partial^2 L_{\mathrm{MD}}}{\partial \theta_{\gamma,n} \partial A_k}\right]_{N_{\boldsymbol{\gamma}}\times K} = \left[\sum_i \frac{1}{\lambda_i}\frac{\partial \gamma_i}{\partial \theta_{\gamma,n}}\frac{\partial \lambda_i}{\partial A_k}\right]_{N_{\boldsymbol{\gamma}}\times K} \tag{A-2}$$

$$\frac{\partial^2 L_{\mathrm{MD}}}{\partial \boldsymbol{\theta}_{\boldsymbol{\lambda}} \partial \mathbf{A}} = \left[\frac{\partial^2 L_{\mathrm{MD}}}{\partial \theta_{\lambda,n} \partial A_k}\right]_{N_{\boldsymbol{\lambda}}\times K} = \left[\sum_i \left(1 - \frac{\gamma_i}{\lambda_i}\right)\frac{\partial^2 \lambda_i}{\partial \theta_{\lambda,n} \partial A_k} - \frac{\gamma_i}{\lambda_i^2}\frac{\partial \lambda_i}{\partial \theta_{\lambda,n}}\frac{\partial \lambda_i}{\partial A_k}\right]_{N_{\boldsymbol{\lambda}}\times K} \tag{A-3}$$

Here, $\frac{\partial^2 L_{\mathrm{MD}}}{\partial \mathbf{A}^2}$, $\frac{\partial^2 L_{\mathrm{MD}}}{\partial \boldsymbol{\theta}_{\boldsymbol{\gamma}} \partial \mathbf{A}}$, and $\frac{\partial^2 L_{\mathrm{MD}}}{\partial \boldsymbol{\theta}_{\boldsymbol{\lambda}} \partial \mathbf{A}}$ are all matrices, with their elements denoted inside the brackets. $K$, $N_{\boldsymbol{\gamma}}$, and $N_{\boldsymbol{\lambda}}$ represent the number of basis materials, parameters in $\boldsymbol{\theta}_{\boldsymbol{\gamma}}$, and parameters in $\boldsymbol{\theta}_{\boldsymbol{\lambda}}$, respectively.

More specifically, for simple models, it is straightforward to hand-code the gradients such as energy threshold correction in our demonstration where $\boldsymbol{\theta}_{\boldsymbol{\lambda}}$ consists of several energy thresholds. For more complicated cases such as the scatter correction network, we chose an intermediate expression, $\partial^2 L_{\mathrm{MD}}/\partial \boldsymbol{\gamma}_c \partial \mathbf{A}$, to calculate $\partial \mathbf{A}^*/\partial \boldsymbol{\gamma}_c$ and rely on the deep learning framework (PyTorch) to manage the gradients of neural network parameters, i.e., $\partial \boldsymbol{\gamma}_c/\partial \boldsymbol{\theta}_{\boldsymbol{\gamma}}$, which together forms $\partial \mathbf{A}^*/\partial \boldsymbol{\theta}_{\boldsymbol{\gamma}}$ for training.

### *B. Differentiable MLE-MD from Unrolled Method*

There are other ways to realize differentiable material decomposition, such as unrolling the iterative solver, which are equivalent in estimating gradients. In this section, we briefly show an interesting observation regarding their GPU footprint differences, which makes the unrolled method impractical for realizing differentiable PCCT. A material decomposition for 100×100×4 pixels of two basis materials (water and calcium) was carried out, and results are shown in Fig. 12. Noticeable differences in GPU usage can be observed between the implicit differentiation-based method and unrolled MD.

Intuitively, as the number of iterations in the iterative solver increases, GPU usage rises linearly in unrolled MD. Considering that each detector may consist of several thousand pixels and each CT scan contains projections from many views, the unrolled method is less practical for making the entire PCCT chain differentiable to enable the end-to-end cross-domain training we aim for.

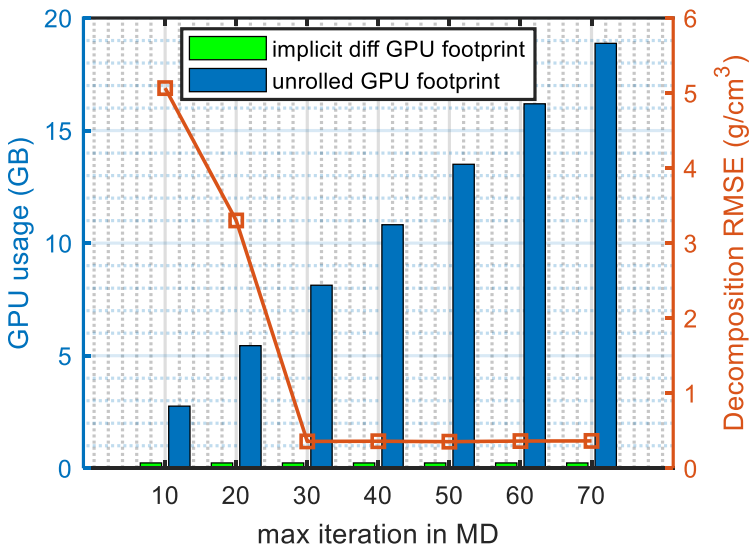

Fig. 12. GPU usage of differentiable MD based on unrolled method and implicit differentiation. Newton's method was used as the solver.

TABLE III
PERFORMANCE OF DIFFERENTIABLE MLE-MD

| | Analytical | | | Proxy |
|---|---|---|---|---|
| Backward | Implicit-Diff | | | Auto-Diff |
| Forward | NM [CPU] | NR [CPU] | NR [GPU] | Proxy MD [GPU] |
| Per Epoch (seconds) | 31000 | 50300 | 406 | 40 |
| Water RMSE (mg/mL) | | | 10.06±3.28 | 9.88±2.87 |
| Calcium RMSE (mg/mL) | \ | | 4.38±2.15 | 3.95±1.83 |
| VMI RMSE (HU) | | | 17.60±8.10 | 17.15±8.22 |

### *C. Differentiable MLE-MD with Analytical Gradients and Proxy Gradients*

The implicit differentiation-based method provides the gradients we need to make the iterative MLE MD differentiable by calculating the gradients analytically at the optimal solution. This also means that in each training step, we need to solve the optimization problem repeatedly. In this section, we briefly demonstrate some techniques we tried for acceleration and the equivalent performance in training among them in TABLE III, including a Nelder-Mead iterative solver (NM), Newton-Raphson iterative solver (NR, both CPU and GPU version), and empirical model-based method (Proxy MD) proposed in our earlier work [32].

For the iterative solver-based method, the forward path outputs the material thickness estimation, and the corresponding gradients are computed from equation (5) in an analytical form. "Analytical" in the table means analytical gradients. As shown in Table III, we implemented several solvers and ultimately selected the GPU version of the Newton–Raphson method to keep the training time within a feasible range. For the Proxy MD method, we used a simple MLP that was trained via derivative-aware Sobolev training. It preserves both the value and the gradients of the iterative solver, thus providing equivalent (proxy) gradients to the analytical gradients, but with much faster execution. Their equivalence in training convergence is demonstrated in the final performance in TABLE III.

## ACKNOWLEDGMENTS

This work was supported by GE HealthCare. This work has been submitted to the IEEE for possible publication. Copyright may be transferred without notice, after which this version may no longer be accessible.